\documentclass[conference, compsoc]{IEEEtran}
\usepackage{graphicx,amssymb,amsmath}
\usepackage{balance}
\balance

\newtheorem{thm}{Theorem}
\newtheorem{defn}[thm]{Definition}
\newcommand*{\metacomp}{\overrightarrow{\cdot}}

\begin{document}

\title{Modeling System Safety Requirements Using Input/Output Constraint Meta-Automata}

\author{\IEEEauthorblockN{Zhe Chen\ $^\dagger$}
\IEEEauthorblockA{$^\dagger$\ Laboratory LATTIS, INSA, University of Toulouse\\
135 Avenue de Rangueil, 31077 Toulouse, France\\
Email: zchen@insa-toulouse.fr} \and \IEEEauthorblockN{Gilles Motet\
$^{\dagger,\ddagger}$}
\IEEEauthorblockA{$^\ddagger$\ Foundation for an Industrial Safety Culture\\
6 All\'ee Emile Monso, 31029 Toulouse, France\\
Email: gilles.motet@insa-toulouse.fr} }

\maketitle

\begin{abstract}
Most recent software related accidents have been system accidents.
To validate the absence of system hazards concerning dysfunctional
interactions, industrials call for approaches of modeling system
safety requirements and interaction constraints among components and
with environments (e.g., between humans and machines). This paper
proposes a framework based on input/output constraint meta-automata,
which restricts system behavior at the meta level. This approach can
formally model safe interactions between a system and its
environment or among its components. This framework differs from the
framework of the traditional model checking. It explicitly separates
the tasks of product engineers and safety engineers, and provides a
top-down technique for modeling a system with safety constraints,
and for automatically composing a safe system that conforms to
safety requirements. The contributions of this work include
formalizing system safety requirements and a way of automatically
ensuring system safety.
\end{abstract}

\IEEEpeerreviewmaketitle

\section{System Accidents}
Computer technology has created a quiet revolution in most fields of
engineering, also introduced new failure modes that are changing the
nature of accidents \cite{Lev04}. To provide much more complex
automated services, industrials have to develop much more
complicated computer systems which consist of numerous components
and a huge number of actions (both internal and interactive). A
recent challenge is the {\em system accident}, caused by increasing
{\em coupling} among system components (software, control system,
electromechanical and human), and their {\em interactive complexity}
\cite{Lev08}\cite{Per99}. In contrast, accidents arising from
component failures are termed {\em component failure accidents}.

{\em System safety} and {\em component reliability} are different.
They are system property and component property, respectively
\cite{Lev08}. {\em Reliability} is defined as the probability that a
component satisfies its specified behavioral requirements, whereas
{\em safety} is defined as the absence of accidents --- events
involving an unacceptable loss \cite{Lev95}. People are now
constructing intellectually unmanageable software systems that go
beyond human cognitive limits, and this allows potentially unsafe
interactions to be undetected. Accidents often result from hazardous
interactions among perfectly functioning components.

There are several examples of system accidents.

The space shuttle {\em Challenger} accident was due to the release
of hot propellant gases from a field joint. An O-ring was used to
control the hazard by sealing a tiny gap in the field joint created
by pressure at ignition. However, the design did not effectively
impose the {\em required constraints} on the propellant gas release
(i.e., it did not adequately seal the gap), leading to an explosion
and the loss of the shuttle and its crew \cite{Lev01}.

The self-destructing explosion of Ariane 5 launcher was resulted
from the successive failures of the active inertial reference system
(IRS) and the backup IRS \cite{Lev01}. Ariane 5 adopted the same
reference system as Ariane 4. However, the profile of Ariane 5 was
different from that of Ariane 4 --- the acceleration communicated as
input value to IRS of Ariane 5 was higher. Furthermore, the {\em
interactions between IRS and other components} were not checked and
redefined. Due to the overflow of input value computation, the IRS
stopped working \cite{KT06}. Then, the signaled error was
interpreted as a launcher attitude, and led the control system to
rotate the tailpipe at the end stop \cite{GM02}.

The loss of the Mars Polar Lander was attributed to a {\em
misapprehensive interaction} between the onboard software and the
landing leg system \cite{JPL00}. The landing leg system was expected
and specified to generate noise (spurious signals) when the landing
legs were deployed during descent. However, the onboard software
interpreted these signals as an indication that landing occurred (as
specified in their requirements) and shut the engines down, causing
the spacecraft to crash into the Mars surface.

\begin{figure}
\centering
  \includegraphics[scale=0.4]{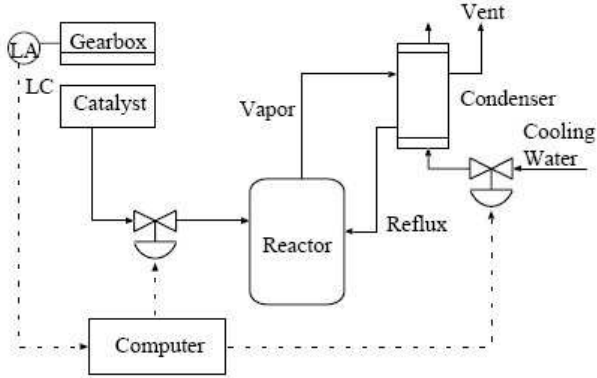}\\
  \caption{A Chemical Reactor Design}\label{Fig:Reactor}
\end{figure}

A system accident occurred in a batch chemical reactor in England
\cite{Kle82}. The computer controlled the input flow of cooling
water into the condenser and the input flow of catalyst into the
reactor by manipulating the valves. The computer was told that if
any component in the plant gets abnormal, it had to leave all
controlled variables as they were and to sound an alarm. On one
occasion, the computer just started to increase the cooling water
flow, after a catalyst had been added into the reactor. Then the
computer received an abnormal signal indicating a low oil level in a
gearbox, and it reacted as its requirements specified: sounded an
alarm and maintained all the control variables with their present
condition. Since the water flow was kept at a low rate, then the
reactor overheated, the relief valve lifted and the contents of the
reactor were discharged into the atmosphere. The design of the
system is shown in Fig. \ref{Fig:Reactor}.

In all these accidents, the components are reliable in terms of
satisfying their specified requirements, but the systems are not
safe as a whole. As Leveson mentioned in STAMP (Systems-Theoretic
Accident Model and Processes) \cite{Lev04}, these accidents result
from inadequate {\em control} or enforcement of {\em safety-related
constraints} of the systems.

Since most software related accidents have been system accidents
\cite{Lev04}, people need to model and constrain interactions of
system components to validate the absence of {\em dysfunctional
interactions}. One of the challenges to system safety is that {\em
the ambiguity in safety requirements} may lead to reliable unsafe
systems. Another challenge is {\em the lack of formal techniques}
for describing safety rules and interactions between components
(including interactions between humans and machines), which makes
automated verifications difficult. In this paper, we will propose a
formal framework for modeling system safety rules. The framework is
based on a new concept of I/O constraint meta-automata.

This paper is organized as follows: the framework of our approach is
proposed in Section 2. The formal technique based on I/O constraint
meta-automata is introduced in Section 3. An example is used to
illustrate how to formalize safety constraints and combine it with a
system specification. In Section 4, we discuss how to apply this
approach to a system that consists of multiple components. In
Section 5, we compare our works to classic verification techniques,
such as model checking, and conclude the paper.

\section{The Framework of Our Approach}
The most popular technique of system safety verification is {\em
model checking} \cite{CGP00}. In this framework, we have two steps
in verifying a system. At first, we formalize system behavior as a
model (e.g., a transition system, a Kripke model \cite{HR04}). At
the second step, we specify the features that we aim at validating,
and use a certain checking algorithm to search for a counterexample
which is an execution trace violating the specified features. If the
algorithm finds such a counterexample, we have to modify the
original system to ensure safety requirements, or else the
verification succeeds.

Unlike the model checking, our framework takes another way. It
consists of the following steps:
\begin{enumerate}
  \item Modeling system behavior, including specifications of its
  components, internal and external interactions.
  \item Modeling system safety constraints using a certain formal technique,
   e.g., I/O constraint meta-automata in this paper.
  \item Combining these two models to deduce a safe system model, that is,
  a system model whose behavior is in accordance with its safety
  constraints.
\end{enumerate}

As we mentioned in \cite{Mot09}, the system behavior specifies an
{\em operational semantics}, which defines what a system is able to
do. Modeling system behavior is mainly performed by {\em product
engineers} (designers), such as programmers and developers. In the
example of the chemical reactor control system, the actions
``opening the catalyst flow'', ``opening the cooling water flow''
and ``sounding an alarm'' are system behaviors.

In the second step, the model of safety requirements specifies a
{\em correctness semantics}, which defines what a system is
authorized to do. This process is the duty of {\em safety engineers}
whose responsibility is to assure system safety. Safety engineers
may consist of requirement engineers, test engineers, managers from
higher socio-technical levels who define safety standards or
regulations \cite{Lev04}, etc. In the example of the chemical
reactor system, the constraint ``opening the catalyst flow must be
followed by opening the cooling water flow'' is an instance of
system safety requirements.

In the third step, in order to ensure system safety, we combine the
system model with its safety constraints model. Then we can check if
the system is safe under the constraints specifying safety
requirements. However, the precondition of such a formal checking is
that we must formalize safety requirements. And we also need to
carefully define the composition of a system model and its
constraints model. In the next section, we will introduce such an
approach based on I/O constraint meta-automata.

We remark here that an other precondition is that we can find all
safety constraints in a system. However, this is an issue of ``risk
identification'' \cite{ISO31000}, which is outside the scope of this
paper. This work deals with ``risk treatment'', which is a different
phase to risk identification, according to the ISO Standard 31000
\cite{ISO31000}.

\section{Modeling System Safety Constraints Using I/O Constraint Meta-Automata}

The theory of input/output automata \cite{LT89}\cite{Lyn96} extends
classic automata theory \cite{HU79} for modeling concurrent systems
with different input, output and internal actions. I/O automata and
the variants are widely used in modeling distributed systems
\cite{Lyn03}.

\begin{defn}
An \textbf{input/output automaton} (also called an I/O automaton or
simply an automaton) is a tuple
$A=(Q,\Sigma^I,\Sigma^O,\Sigma^H,\delta,S)$, where:

\begin{itemize}
  \item $Q$ is a set of \textbf{states}.
  \item $\Sigma^I,\Sigma^O,\Sigma^H$ are
pairwise disjoint sets of \textbf{input, output and internal
actions}, respectively. Let $\Sigma = \Sigma^I \bigcup \Sigma^O
\bigcup \Sigma^H$ be the set of \textbf{actions}.
  \item $\delta \subseteq
Q \times \Sigma \times Q$ is a set of \textbf{labeled transitions},
such that for each $a \in \Sigma^I$ and $q \in Q$ there is a
transition $p_k:(q,a,q') \in \delta$ (\textbf{input-enabled}).
  \item $S \subseteq Q$ is a nonempty set of \textbf{start states}. \hfill $\Box$
\end{itemize}
\end{defn}

In the graph notation, a transition $p_k:(q,a,q')$ is denoted by an
arc from $q$ to $q'$ labeled $p_k:a$, where $p_k$ is the name of the
transition. To discriminate explicitly the different sets of actions
in diagrams, we may suffix a symbol ``?'', ``!'' or ``;'' to an
input, output or internal action, respectively.

In the example of the batch chemical reactor, the computer system
behavior is modeled using an I/O automaton $A$ of Fig.
\ref{Fig:DFA_RCS}(1). The automaton $A$ includes a set of input
actions $\Sigma^I = \{l\}$ (low oil signal), a set of output actions
$\Sigma^O = \{c,w,a\}$ (opening catalyst flow, opening water flow,
sounding an alarm, respectively), and a set of internal actions
$\Sigma^H=\{ e \}$ (ending all operations). The normal operational
behavior includes opening catalyst flow ($p_1$), then opening water
flow ($p_2$), etc., resulting in an infinite execution trace
$p_1p_2p_1p_2...$. To respond to abnormal signals as soon as
possible, all the states have a transition labeled $l$, which leads
to a state that can sound an alarm ($p_6$) and stop the process
($p_8$). Unfortunately, this design leads to hazardous behaviors:
$(cw)^* clae$, that is, after a sequence of opening catalyst and
water flows $(cw)^*$, then the catalyst flow is opened $c$ when an
abnormal signal is received $l$, then an alarm is sounded $a$. So
water is not added after the catalyst flow is opened. This sequence
of events leads to the accident mentioned in Section 1.

\begin{figure*}
  \centering
  \includegraphics{fig_mss.1}\ \ \ \ \ \includegraphics{fig_mss.2}\\
  \caption{Automata of the Reactor Control System}\label{Fig:DFA_RCS}
\end{figure*}

Note that this hazard is due to the uncontrolled sequences of
transitions --- $p_1$ must be followed by $p_2$ and not by $p_4$. To
solve this problem, we need to specify the authorized sequences
(satisfying safety constraints) on the transitions $\delta$ and not
on the actions $\Sigma$. Thus, these constraints are not at the
behavioral model level, but at the meta-model level. We propose the
concept of {\em constraint meta-automata} to formalize safety
constraints. Then, we combine a meta-automaton with the system
automaton.

\begin{defn}
A \textbf{constraint meta-automaton} (or simply meta-automaton)
$\hat{A}$ over an I/O automaton $A=(Q,\Sigma,\delta,S)$ is a tuple
$\hat{A}=(\hat{Q},\hat{\Sigma},\hat{\delta},\hat{S})$, where:

\begin{itemize}
  \item $\hat{Q}$ is a set of \textbf{states} disjoint with $Q$.
  \item $\hat{\Sigma}$ is a set of \textbf{terminals} that consists of all
the transition names in $\delta$ of $A$.
  \item $\hat{\delta}$ is a set of \textbf{labeled transitions}.
  \item $\hat{S} \subseteq \hat{Q}$ is a nonempty set of \textbf{start
states}. \hfill $\Box$
\end{itemize}
\end{defn}

\begin{figure}[htb]
  \centering
  \includegraphics{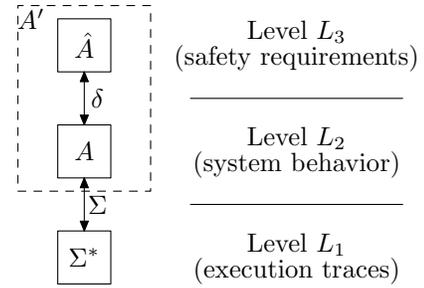}\\
  \caption{A 3-levels Overview}\label{Fig:3_levels}
\end{figure}

Note that the transitions $\delta$ of $A$ are terminals of
$\hat{A}$, so we say that $\hat{A}$ is at the meta level of $A$.
Figure \ref{Fig:3_levels} illustrates the 3 levels in our framework.
Let $\Sigma^*$ be a set of execution traces of actions, $A$
describes the behavior on $\Sigma$. $\hat{A}$ specifies the behavior
on the $A$-transitions ($\hat{\Sigma}=\delta$), that is, a behavior
on the behavior of $A$. This meta-behavior expresses safety
requirements.

In the example, to prevent accidents, we need to bind the safety
constraint ``opening catalyst must be followed by opening water,''
that is, ``whenever the transition $p_1:c$ occurs, the transition
$p_2:w$ must occur after that.'' This constraint can be formalized
as a constraint meta-automaton $\hat{A}$ of Fig.
\ref{Fig:DFA_RCS}(2). When we design this constraint, we only
specify the sequence of transitions $p_1,p_2$ at the meta-model
level, and we concern little about the implementation of the system
at the model level. The next step is to compose the system automaton
$A$ with its constraint meta-automaton $\hat{A}$, and automatically
generate a system model $A'$ satisfying the safety requirement.

\begin{defn}
The \textbf{meta-composition} $A'$ of an I/O automaton
$A=(Q,\Sigma,\delta,S)$ and a constraint meta-automaton
$\hat{A}=(\hat{Q},\hat{\Sigma},\hat{\delta},\hat{S})$ over $A$ is a
tuple:
\begin{equation}
A' = A \metacomp \hat{A} = (Q \times \hat{Q}, \Sigma, \delta', S
\times \hat{S})
\end{equation}
where $p_k: ((q_i,\hat{q}_j), a, (q_m,\hat{q}_n)) \in \delta'$ iff,

(1) $p_k:(q_i,a,q_m) \in \delta$, and

(2) $(\hat{q}_j, p_k, \hat{q}_n) \in \hat{\delta}$. \hfill $\Box$
\end{defn}

The symbol $\metacomp$ is the {\em meta-composition operator}, and
read as ``meta-compose''. Its left and right operands are an
automaton and a constraint meta-automaton, respectively.

\begin{figure}[htb]
  \centering
  \includegraphics{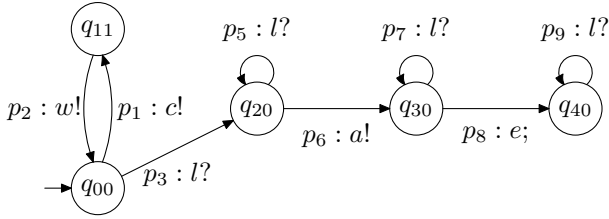}\\
  \caption{The Meta-Composition $A'$ } \label{Fig:DFA_RCS_C}
\end{figure}

Notice that $\delta = \{p_k\}_{k \in \mathcal{K}}$ plays a key role
in associating transitions of $A$ and terminals of $\hat{A}$. For
our example, we combine the automata $A$ and $\hat{A}$ of Fig.
\ref{Fig:DFA_RCS}, thus we get the automaton $A' = A \metacomp
\hat{A}$ of Fig. \ref{Fig:DFA_RCS_C} where $q_{ij}$ denotes
$(q_i,\hat{q}_j)$.

The meta-composition contains exactly all the paths satisfying the
constraint in the system. Formally, we have the following theorem
(the proof is omitted for its simpleness and intuitiveness from the
definition):

\begin{thm}
Given $A,\hat{A}$ and the meta-composition $A'$, an
\textbf{execution trace} $t_\Sigma \in \Sigma^*$ is in $A'$ iff,
$t_\Sigma$ is in $A$, and its \textbf{transition trace} $t_\delta
\in \delta^*$ is in $\hat{A}$. \hfill $\Box$
\end{thm}

Obviously, the set of traces of $A'$ is a subset of the traces of
$A$. Formally, let $L(A)$ be the set of traces of $A$ (also the
language of $A$), we have $L(A') \subseteq L(A)$.

Thanks to $\hat{A}$, the hazardous execution traces, for example
$cwclae$, which exists in $A$, will be eliminated, because its
transition trace $p_1p_2p_1p_4p_6p_8 \not\in L(\hat{A})$ (the
language of $\hat{A}$). The comparison between $A$ of Fig.
\ref{Fig:DFA_RCS}(1) and $A'$ of Fig. \ref{Fig:DFA_RCS_C} highlights
the hazardous transition $p_4$ of $A$. However, in general, this
diagnosis is much more complex and cannot be achieved manually,
since a real system $A$ has too many states to be expressed clearly
on a paper. That is why we should provide a formal and automated
method for eliminating hazardous transitions.

\section{Modeling Multi-Component Systems with Safety Constraints}

Our approach can also be applied to the systems that are made up of
several components, whose safety constraints are related to several
components.

As a preliminary, we redefine the composition of I/O automata, which
was introduced in \cite{LT89}.

Let $\mathcal{N} = \{n_1,...,n_k\} \subseteq \mathbb{N}$ be a finite
set with cardinality $k$, and for each $n \in \mathcal{N}$, $S_n$ be
a set. Then we define: $\prod_{n \in \mathcal{N}}S_n
\stackrel{\text{def}}{=} \{ (x_{n_1},x_{n_2},...,x_{n_k})\ |\
(\forall j \in \{1,...,k\} \bullet x_{n_j} \in S_{n_j}) \wedge
\mathcal{N} = \{n_1,...,n_k\} \wedge (\forall j_1,j_2 \in
\{1,...,k\} \bullet j_1<j_2 \rightarrow n_{j_1}<n_{j_2}) \}$. We
define the function of projection $\overrightarrow{s}[j]$ to denote
the $j$-th component of the state vector $\overrightarrow{s}$:
$\forall j \in \{1,...,k\}$, $(x_{n_1},x_{n_2},...,x_{n_k})[j]$ =
$x_{n_j}$.

\begin{defn}
A finite collection of I/O automata $\{A_n\}_{n \in \mathcal{N}}$ is
said to be \textbf{strongly compatible} if $\forall i,j \in
\mathcal{N}$, $i \neq j$, we have

(1) $\Sigma^O_i \cap \Sigma^O_j = \emptyset$, and

(2) $\Sigma^H_i \cap \Sigma_j = \emptyset$. \hfill $\Box$
\end{defn}

\begin{defn}
The \textbf{composition} $A = \prod_{n \in \mathcal{N}} A_n$ of a
finite collection of strongly compatible I/O automata $\{A_n\}_{n
\in \mathcal{N}}$ is an I/O automaton $(\prod_{n \in
\mathcal{N}}Q_n, \Sigma^I, \Sigma^O, \Sigma^H, \delta, \prod_{n \in
\mathcal{N}}S_n)$ iff,

\begin{itemize}
  \item $\Sigma^I = \bigcup_{n \in \mathcal{N}}\Sigma^I_n - \bigcup_{n \in
\mathcal{N}}\Sigma^O_n$,
  \item $\Sigma^O = \bigcup_{n \in \mathcal{N}}\Sigma^O_n$,
  \item $\Sigma^H = \bigcup_{n \in \mathcal{N}}\Sigma^H_n$, and
  \item for each $\overrightarrow{q}, \overrightarrow{q}' \in \prod_{n \in \mathcal{N}}Q_n$ and $a \in
\Sigma$, \\
$p_\mathcal{I} : (\overrightarrow{q}, a,
\overrightarrow{q}')\in \delta$ iff $\forall j: 1 \leq j \leq
|\mathcal{N}| \wedge {n_j \in \mathcal{N}}$,

\begin{enumerate}
  \item if $a \in \Sigma_{n_j}$ then \\
$\exists i: i \subseteq \mathcal{I} \bullet p_i:
(\overrightarrow{q}[j], a, \overrightarrow{q}'[j]) \in \delta_{n_j}
$;
  \item if $a \not\in \Sigma_{n_j}$ then $\overrightarrow{q}[j] =
\overrightarrow{q}'[j]$ and \\
$\forall i: p_i \in \delta_{n_j} \bullet i \cap \mathcal{I} =
\emptyset$. \hfill $\Box$
\end{enumerate}

\end{itemize}
\end{defn}

Notice that the name of a transition of $A$ may contain a set of
names of original transitions $p_{\mathcal{I}} = \{ p_i \}_{i
\subseteq \mathcal{I}}$, where $i$ may be a set or a single element.

\begin{figure*}
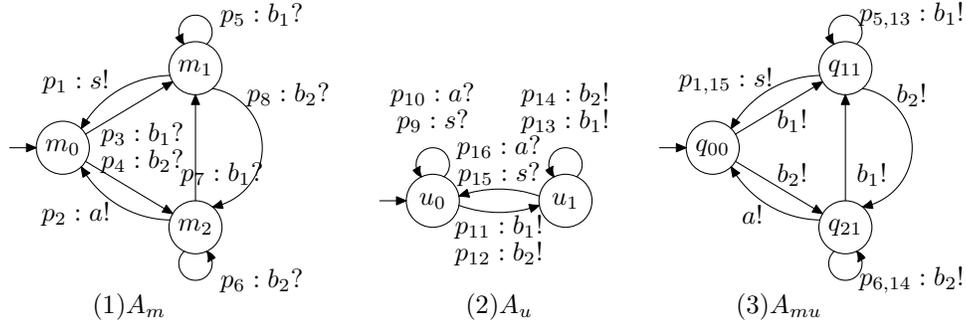

  \centering
  \includegraphics{fig_mss.4}\ \ \ \ \ \includegraphics{fig_mss.5}\ \ \ \ \ \includegraphics{fig_mss.6}\\
  \caption{Automata of the Candy Machine System}\label{Fig:DFA_CM}
\end{figure*}

We use an example derived from \cite{LT89}, concerning a system
composed of two components with interactions: a candy vending
machine and a customer. The candy machine $A_m$, specified in Fig.
\ref{Fig:DFA_CM}(1), may receive inputs $b_1,b_2$ indicating that
buttons 1 and 2 are pushed, respectively. It may output $s,a$,
indicating candy dispensation actions, SKYBARs and ALMONDJOYs,
respectively. The machine may receive several inputs before
delivering a candy. A greedy user $A_u$, specified in Fig.
\ref{Fig:DFA_CM}(2), can push buttons $b_1,b_2$ or get a candy
$s,a$. The greedy user does not wait for a candy bar before pressing
a button again.

The composition of the machine behavior and the user behavior is
defined by $A_{mu}=A_m \cdot A_u$ of Fig. \ref{Fig:DFA_CM}(3), where
$q_{ij}$ denotes the composite state $(m_i,u_j)$, $p_{i_1, \ldots
,i_k}$ denotes a set of transitions $\{ p_{i_1}, p_{i_2}, \ldots ,
p_{i_k} \}$. A transition of the composition may be composed of
several transitions of components. For example, $p_{1,15}:s$ is a
synchronization of $p_1:s!$ and $p_{15}:s?$, which belong to $A_m$
and $A_u$, respectively. Formally, a transition of $A = \prod_{n \in
\mathcal{N}} A_n$ may be composed of $i$ transitions of components,
where $1\leq i \leq |\mathcal{N}|$.

In the context of composite transitions, a composite transition is
allowed iff one of its sub-transitions is authorized by its
constraint meta-automaton. Thus, we define the meta-composition
operator as follows:

\begin{defn}
The \textbf{meta-composition} $A'$ of a composition $A = \prod_{n
\in \mathcal{N}} A_n = (\prod_{n \in \mathcal{N}} Q_n
,\Sigma,\delta, \prod_{n \in \mathcal{N}} S_n)$ and a constraint
meta-automaton $\hat{A}=(\hat{Q},\hat{\Sigma},\hat{\delta},\hat{S})$
over $A$ is a tuple:
\begin{equation}
A' = A \metacomp \hat{A} = ( (\prod_{n \in \mathcal{N}} Q_n) \times
\hat{Q}, \Sigma, \delta', (\prod_{n \in \mathcal{N}} S_n) \times
\hat{S})
\end{equation}
where $p_{\mathcal{I}}: ((\overrightarrow{q_i}, \hat{q}_j), a,
(\overrightarrow{q_m}, \hat{q}_n)) \in \delta'$ iff,

(1) $p_{\mathcal{I}}: (\overrightarrow{q_i}, a,
\overrightarrow{q_m}) \in \delta$, and

(2) $\exists k: k \in \mathcal{I} \bullet (\hat{q}_j, p_k,
\hat{q}_n) \in \hat{\delta}$. \hfill $\Box$
\end{defn}

Notice that the specification of the example allows a hazardous
situation: the greedy user repeatedly pushes a single button without
giving the machine a chance to dispense a candy bar (the transition
labeled $p_{5,13}:b_1$ of $q_{11}$ does not allow the transition
$(q_{11}, s, q_{00})$ to be fired). To prevent this situation, the
following constraints forbid successive occurrences of pressing a
single button:
\begin{itemize}
  \item Whenever one of the transitions $p_3,p_5,p_7$ (action
$b_1$) occurs, the next transition must be not $p_3,p_5,p_7$.
  \item Whenever one of the transitions $p_4,p_6,p_8$ (action
$b_2$) occurs, the next transition must be not $p_4,p_6,p_8$.
\end{itemize}

Differing from the previous example, the constraint needs to
synchronize the actions of the machine and of the user.

\begin{figure*}
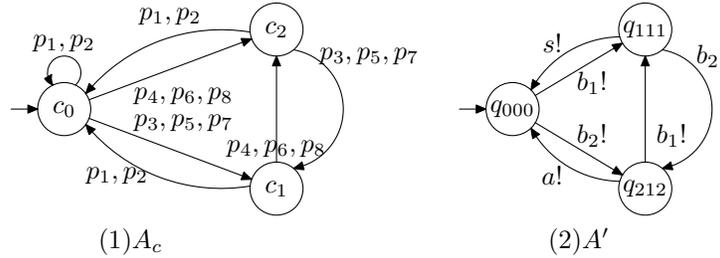

  \centering
  \includegraphics{fig_mss.7}\ \ \ \ \includegraphics{fig_mss.8}\\
  \caption{A Safety Constraint of the Candy Machine System}\label{Fig:DFA_CM_C}
\end{figure*}

Formalizing the constraints, the semantics of the constraint
meta-automaton $A_c$ of Fig. \ref{Fig:DFA_CM_C}(1) is: whenever the
user pushes a button, she or he cannot push it again, but may push
the other button to change the choice, or wait for a candy bar.

Combining the whole system $A_{mu}$ with its constraint $A_c$, we
get the system $A' = (A_m \cdot A_u) \metacomp A_c$ in Fig.
\ref{Fig:DFA_CM_C}(2), where $q_{ijk}$ denotes the composite state
$(m_i,u_j,c_k)$. All of its execution traces satisfy the constraint,
and thus prevent the hazardous situation.

This simple example is good for demonstrating the principle,
avoiding indigestible diagrams of automata. Since we formally
defined the {\em meta-composition operator}, it can be easily
implemented to be an automated tool. Thus, it can be applied to more
complex systems.

\section{Conclusion}
We propose modeling system safety requirements formally using I/O
constraint meta-automata. As we illustrated using the examples, this
approach can formally model safe interactions between a system and
its environments, or among its components. This framework differs
from the one of the traditional model checking. It explicitly
separates the tasks of product engineers and safety engineers, and
provides a technique for modeling a system with safety constraints,
and for automatically composing a safe system that conforms to
safety requirements.

The essential ideas of our approach are the separation and
formalization of the system specification $A$ (behavioral
requirements) and the safety constraints $\hat{A}$ (safety
requirements). The automaton $A$ handles inputs to produce outputs
using activities depending on the states, whereas the meta-automaton
$\hat{A}$ treats activities to produce the set of acceptable
activities depending on safety requirements.

Our framework has different objective and uses different approaches
to those of model checking. Model checking techniques use {\em a
bottom-up approach} --- it verifies execution traces $\Sigma^*$ at
the lower level $L_1$ to prove the correctness and safety of the
system model $A$ at the middle level $L_2$ (see Fig.
\ref{Fig:3_levels}). However, our proposal uses {\em a top-down
approach} --- we model safety requirements as acceptable sequences
of transitions ($\delta^*$) at the higher level $L_3$ to ensure the
correct use of $A$. Then any execution trace (at $L_1$) that
conforms to the meta-composition $A'$ is definitely a safe
execution. So the two techniques are complementary. Model checking
may be used to reduce the fault likelihood, and our approach can be
applied to avoid behavior that are not in accordance with some
critical safety requirements.

Both linear time logic and branching time logic have been proved to
be useful in checking properties of traces of classic automata
\cite{CGP00}. Since I/O automata are extended from classic automata,
these existing techniques can be easily applied to I/O automata with
little modifications.

In the future, we will apply the approach to the variants of I/O
automata, e.g., timed, hybrid, probabilistic, dynamic \cite{Lyn03}.
Our approach can also be applied to the systems specified using
classic automata \cite{HU79}, since I/O automata are specific
extensions of traditional automata.

We will also study the formalization of parameterized constraints.
To model parameterized systems more accurately, parameters of
actions and value domains of variables should be considered. This is
also the basis of studying reusability, substitutability and
equivalence of components.

As we mentioned, identification of potential hazards is also a
challenge in practice. Risk identification and treatment are both
important phases in risk management \cite{ISO31000}. This will be a
good direction for future work.

\bibliographystyle{IEEEtran}
\bibliography{main}

\end{document}